\documentclass[onecolumn, prl, showkeys, nofootinbib, letterpaper, dvips]{revtex4}

\usepackage[%
    dvips,%
    pdftitle = {{Exam fairness}},%
	pdfsubject = {{It is widely agreed that exams must be fair; yet what this exactly means is not made clear. One may mean fairness of treatment, but this merely propagates the fairness or unfairness of pre-existing rules. Fairness of opportunity on the other hand necessarily leads to identical grades for everyone, which clearly makes it inapplicable. Neither view is helpful to make decisions on competing claims: fairness of treatment ignores the problem and fairness of opportunity holds all claims to be equally valid. To escape this deadlock one needs an external criterion, such as how good engineers students will be, to replace fairness viewed as student--student comparison.}},%
	pdfauthor = {{Mathieu Bouville}},%
	pdfkeywords = {{assessment; bias; engineering education; examinations; grading; justice; tests}},%
    hyperindex,%
    bookmarksopen,%
    bookmarksopenlevel=1,%
]{hyperref}

\begin{document}

\title{Exam fairness}
\addcontentsline{toc}{chapter}{EXAM FAIRNESS -- M. BOUVILLE}

\author{Mathieu Bouville}
	\email{m-bouville@imre.a-star.edu.sg}
	\affiliation{Institute of Materials Research and Engineering, A*STAR, 
Singapore 117602}
	\affiliation{Institute of High Performance Computing, A*STAR, Singapore 117528}

\begin{abstract}
It is widely agreed that exams must be fair; yet what this exactly means is not made clear. One may mean fairness of treatment, but this merely propagates the fairness or unfairness of pre-existing rules. Fairness of opportunity on the other hand necessarily leads to identical grades for everyone, which clearly makes it inapplicable. Neither view is helpful to make decisions on competing claims: fairness of treatment ignores the problem and fairness of opportunity holds all claims to be equally valid. To escape this deadlock one needs an external criterion, such as how good engineers students will be, to replace fairness viewed as student--student comparison.
\hspace{\fill}\copyright~Mathieu Bouville, 2008
\end{abstract}
\maketitle

\vspace{2mm}

\begin{quote}
``What students hate more than anything else are examinations that they perceive as unfair''~\cite{Felder-tests}
\end{quote}

\begin{quote}
``For the student, the most important question to be answered in order to be content with the outcomes of the assessment is `what is a fair assessment?'\,'' \cite{Vos-EJEE-00}
\end{quote}

\begin{quote}
``85.1 percent of the sample male, and 97.5\% of the female sample students indicated `definitely', `yes', and `to some extent', to the positive influence of grading fairness in their evaluation of faculty'' \cite{Koushki-IJEE-05}
\end{quote}

There is a wide-spread consensus ---amongst both teachers and students--- that exams must be fair. However, what this exactly means is unclear: even though one can see what `fair' may mean in the case of grading, what it means for the exam itself to be fair is not obvious. 
If one were to ask around what a fair test would be like, one would receive many answers, some of which mutually exclusive. (Imagine that an American and a Briton agree that `chips taste good'. Since the word `chips' does not have the same meaning for the two of them, this agreement is a semantic artifact: they refer to different things for which they simply happen to use the same word~\cite{Bouville-diversity}.) If people do not agree on what `fair' means then the consensus that `exams must be fair' is illusory. And if one does not know what it means for an exam to be fair one cannot design fair exams. In fact, \citet{Green-07} surveyed secondary teachers and found that ``assessment is a realm without professional consensus.'' 
The purpose of this article is to ask explicitly what it means for an exam to be fair. Several possible conceptions of fairness will be presented and analyzed in the context of examinations.

\section{Exam fairness: Possible interpretations}

According to \citet{Felder-tests}, students deem an exam unfair in the following cases (\citet{Vos-EJEE-00} gives similar~criteria):
\begin{quote}
(1)~problems on content not covered in lectures or homework assignments; 
(2)~problems the students consider tricky, with unfamiliar twists that must be worked out on the spur of the moment; 
(3)~excessive length, so that only the best students can finish in the allotted time; 
(4)~excessively harsh grading, with little distinction being made between major conceptual errors and minor calculation mistakes; 
(5)~inconsistent grading, so that two students who make the identical mistake lose different points. 
\end{quote}
\noindent While the first four issues may be indicative of weaknesses of the exam, they do not obviously qualify as~unfair. (In fact, `flaws'~1 and~2 tend to reduce the grade of all students rather than favor some over others.) 
It is rather common for the word `unfair' yo have such a vague negative connotation~--- `this exam was unfair' standing for `there was something wrong with this exam.' (There can be a further drift, with `unfair' losing any objectivity: `this exam was unfair' is sometimes used instead of `I do not like my grade.' While it is true that ``unfair and poorly graded exams cause student resentment''~\cite[p.~214]{Wankat-Oreovicz}, this does not entail that resentment can be equated with unfairness~--- a good grade, however unfair, will not make a student complain.) This article does not mean to study perceived unfairness nor what generally makes a good exam; `fairness' will thus be used in a rather specific sense.

According to the Committee on the Foundations of Assessment of the National Research Council (hereafter~`NRC'),
\begin{quote}
The inferences drawn from an assessment [must be] \emph{valid}, \emph{reliable}, and \emph{fair}. [\,\ldots] Fairness en\-com\-passes a broad range of interconnected issues, including absence of bias in the assessment tasks, equitable treatment of all examinees in the assessment process, opportunity to learn the material being assessed, and comparable validity (if the test scores underestimate or overestimate the competencies of members of particular groups, the assessment is considered unfair).~\cite[p.~39]{NRC-01}
\end{quote}
\noindent One should note that, of the three criteria, `fairness' is the least precise: (i)~it is a non-exhaustive list; (ii)~it relies on concepts, such as `bias' and `equitable', that are not defined (for instance, it is unclear whether ``absence of bias'' and ``equitable treatment'' are different); (iii)~it otherwise assumes knowledge that may not exist: in order to know whether one `overestimate[s] the competencies' one needs to know independently what these competencies are, which may require another test (which would have to meet the same criterion of fairness); and (iv)~it does not specify what these ``particular groups'' are or if there is a rule to determine what they are.
This definition is therefore unsatisfactory. In order to clarify the issue, this article will introduce several possible conceptions of fairness and consider whether they are suitable to define a `fair~exam'.%
\footnote{Exams may have several purposes: they may be used to assign a grade, to give feedback to the students, to let the teacher know about what the students do not understand, etc. For the latter two, there is no question of fairness: the question is whether the exam provides the relevant information. Moreover, fairness is a concern of students only when grades are involved: one is unlikely to hear a student complain `this exam was unfair because I didn't receive enough feedback.' One should also note that a great deal of feedback is done through the grade itself (one may not like this state of affairs but in fact the grade is often the only feedback the students receive). This article will focus on the role of exams having to do with grading and ranking.}

\section{Fairness of treatment}
By fairness, one may mean fairness of treatment: the same rule is applied to all students; for instance, the same answer should get the same points. In particular, one may say that an exam is fair if it is based on explicit class objectives~--- say what you do and do what you say. The ability to solve simple problems quickly is then an acceptable criterion if it was an explicit objective of the class. But what if some students complain that such an objective is unfair because it favors students who are naturally fast or because students whose strength is their ability to cope with complex issues are at a disadvantage? One will reply that this ability is not part of the objectives, and the student will ask why. Saying that we tested on the objectives, that objectives were clearly defined, and that we taught students what would be on the test will not convince them. This process is fair in the sense that tests are based on the objectives; but are objectives them\-selves~fair? 

Let us imagine that we are trying to organize a match-up between Tiger Woods (best golf player in the world) and Roger Federer, best tennis player in the world. If they play golf and the same rules apply to both players Woods will win. But is this really equality of treatment? After all, choosing golf in itself gives an overwhelming edge to Woods. This match is unfair because it is golf. The rules may be the same for the two players, they are still the rules of the sport of one of them. 

Giving credit to students because their name starts with a B is not any less fair than giving them credit because they have the correct answer. In both cases the same rule is applied to all students identically, but it happens to be more convenient for some students than for others (paraphrasing Orwell, one could say that \emph{all students are treated fairly but some are treated more fairly than others}). Fairness of treatment is no fairness at all when it means applying rules that are themselves unfair: it merely propagates the fairness or unfairness of pre-existing rules. It may be relevant to grading but it is not applicable to exam design as a whole. (Those, such as \citet{Bloom-81}, who equate fairness with fairness of grading therefore cannot be right.) Teaching and testing based on class objectives is not sufficient, since the objectives themselves may be unfair. As Ellett~\cite[p.~104]{Ellett-77} notes, whether a procedure is fair ``depends upon the fairness of the results which are brought about by following the procedure or guidelines.'' Saying ``Assess As Ye Would Be Assessed''~\cite{Payne-03} is of no help~--- even though I may personally like the idea that those whose name starts with a~B should be advantaged, this is not an acceptable policy. One needs a richer concept, one that could generate these objectives.

\section{Fairness of opportunity}
A difference between one's name and one's answer is that students do not deserve anything for the former. Perhaps, one should say that ``fair test design should provide examinees comparable opportunity, insofar as possible, to demonstrate knowledge and skills they have acquired that are relevant to the purpose of the test'' \cite[p.~10]{Willingham-97}. Exams should not ``underestimate or overestimate the competencies of members of particular groups'' \cite[][p.~39]{NRC-01} and ``an assessment task is considered biased if construct-irrelevant characteristics of the task result in different meanings for different subgroups'' \cite[][p.~214]{NRC-01}. 

\Citet{Everett-99} ``value students with a diverse set of academic talents'' and do not want to give advantages to those with one talent rather than another. What if a student claims that his specific talent was not taken into account, so that his competences were underestimated and the exam is unfair? 
Coming back to Woods and Federer, one may wonder how to design a match that would be fair in this respect, i.e.\ a match that would account for their very different talents. In any case, if one succeeds in creating such a fair match-up then the outcome must be a tie. Otherwise, the loser would claim that the match was unfair (and there would be no way to prove him wrong). Fairness of opportunity must lead to identical grades, lest losers claim that the exam was unfair to them.

But then a match between either athlete and me should also be a tie. Does it mean that I am as great an athlete as Tiger Woods or Roger Federer? Not at all. It simply means that one cannot assess both the fairness of the match and which player is greater (one cannot solve for two variables based on a single equation). The difficulty springs from the number of possible talents and their incommensurability. In other words, how to rank people who are better in some respects and weaker in others? If there is a single thing which I can do better than Woods then I can argue that his victory is unfair as it undervalues my talent. One may reply that his victory at golf trumps my victory at bowling because of the intrinsic superiority of golf; but how to justify this hierarchy? One may also say that he won in a landslide at golf and I could barely beat him at bowling; but this does not compare him to me, it compares both of us to other players and infers a relative ranking between us. In both cases, one needs external input. His victory is justified only if these external criteria are.

\Citet{NRC-01} assumes that what the ``construct-irrelevant characteristics'' are is obvious and that one can always avoid them. But I just pointed out that this assumption does not hold. Such a view is inapplicable because it provides no criterion for deciding what characteristics are relevant or to rank conflicting relevant characteristics (see also next section). Following \citet{Linn-84}, \citet{Cole-01} note that ``the models of fairness in selection assume that there is a reliable and valid criterion to be predicted that is in itself unbiased. Such criteria very rarely exist.'' I will introduce one such possible criterion in the `Better engineers'~section.

\section{An example: Exam length}
All authors recommend that students be given a lot of time, or even that ``it is best to give tests where there is effectively no time limit'' \cite[p.~221]{Wankat-Oreovicz}. McKeachie~\cite[p.~80]{McKeachie} too ``prefer[s] to give tests without a time limit'' and in any case ``even the slower students [should] have time to finish.'' In some cases, ``only a handful of students have time to finish [the exam], and some who really understand the material fail miserably because the only way they're capable of working is slowly and methodically'' \cite{Felder-tests}. 

It is obvious that ---whatever the length of the exam--- all students are treated in an identical manner, so that fairness of treatment is of no use to decide whether tests should be short or long. On the other hand, a short test (timewise) tends to penalize more conscientious students while a longer exam may penalize faster students. Whatever we do, we are bound to give an advantage to someone: all we can do is choose to whom. Consequently, principles such as ``do not harm''~\cite{Green-07} are of no use to make decisions. 
One can also note that the situation is quite symmetric: some who really understand the material may fail because they are too slow to finish and others may fail because of careless errors. Why then give slower students an edge? Giving students an infinite amount of time would mean that speed is utterly worthless~--- this is plainly~false.

As was pointed out earlier, if careful and fast students do not receive the same grade some may complain about the length of the exam. This means that, if the exam is to be fair, the average grade of conscientious students should be the same as that of fast students. But of course it may be that the conscientious are ---as such--- intrinsically better in this kind of class (e.g.\ labs) or perhaps the conscientious students of this class are better than the fast students of this class. Giving the same average grade to both groups is then unfair since it inflates the grade of the fast students just because they are in the fast group (or it inflates the grade of the more careful students, as the case may be).

There may also be complaints that the grading scheme (in particular the policy regarding partial credit for small mistakes, as mentioned by Felder as criterion 4) favors accuracy too much or on the contrary does not value it enough, so that some students are unfairly advantaged (if merely getting the solution of a problem on track gets nearly full credit then those students who found the correct result are undervalued; if, on the other hand, there is no partial credit then students who are a bit careless will be at a disadvantage). Moreover, since there are dichotomies other than the careful--fast one, other groups may complain about another aspect of the exam so that one will end up giving the same grade to all~--- doing otherwise would be~unfair.

This depicts a somewhat gloomy situation.%
\footnote{One should remark that no exam can be perfectly fair. Imagine that student~D spent days doing problems from various textbooks to prepare for the exam while student~L did only one such problem. Imagine further that the problem L did is on the exam while none of those done by D is. This is unfair --- student~D would have \emph{deserved} the good grade that L received out of \emph{luck}. Yet, nobody can be blamed: the instructor did not favor~L, luck did. In such a case, D~complaining that the exam was unfair is both meaningful and pointless. There are always situations in which someone receives an undeserved advantage without anyone doing anything wrong~\cite{Bouville-cheating}. Part of the unfairness of exams is inexpugnable.}
The most natural conception of fairness is simply treating everyone in the same manner. However, this fairness of treatment merely propagates the fairness or unfairness of rules; it is useless except in the context of grading. Fair opportunity on the other hand necessarily leads to equal results for everyone, which is plainly inapplicable. The example of exam length clearly showed that neither view of fairness is able to generate meaningful criteria to design exams~--- fairness of treatment ignores the problem and fairness of opportunity holds all claims to be equally valid. 
One needs a third view of fairness, which will enable to make concrete decisions regarding exams.

\section{Need for an external criterion}
It is noteworthy that both criteria of fairness examined so far are relative, i.e.\ they compare students with one another: the same treatment relative to other students, the same opportunity relative to other students. This is also true of the aforequoted NRC definition. The problem is that if students have different abilities one cannot tell which student deserves a better grade unless one can tell which ability is more valuable. In other words, it is not possible to simply compare one student to another: one must assess all of them on some external scale. 
This will allow one to discriminate between various talents so that one can meaningfully give different grades to students with different competences. Speed, cautiousness, intelligence, and physical strength all qualify as abilities, but they may not be equally valuable. What is an important ability depends on the context: being tall is good for basketball players while being short is good in the case of jockeys. 
Nobody would call engineering exams unfair because they have engineering questions rather than grammar questions, even though this obviously favors students who are good at engineering against literature majors. It is nevertheless clear that these two competences ---engineering and grammatical--- are not equally valuable in this context. A difference so far overlooked between the Woods--Federer match-up and engineering classes is that the latter have a specific context.

\Citet{Everett-99} insist that ``in disciplines where a variety of skills is valuable, the grading scheme should reward students for demonstrating competency in each of the different skills.'' This section introduces `what would make someone a better engineer' as a possible criterion to establish the relevance of the competing talents. What would make students better engineers seems a meaningful criterion; for instance, better engineers can contribute more to society. They should therefore get a better grade on the exam, a better grade for the course, and a better job.  While accounting for the needs of students as future engineers is not in itself a new idea, my point here is that such a criterion is \emph{necessary} in order for exams to be fair (unless one accepts that all students should get the same grade)~--- an exam which does not account for what would make students better engineers is bound to be unfair.  The best length for exams is then that which corresponds to whatever is the best test of what makes the best engineers. How long this precisely is is an empirical rather than a theoretical question: the balance between speed and accuracy must be based on actual engineering practice.%
\footnote{Naturally, what engineers actually need may depend on the field, the kind of position (e.g.\ R \& D \emph{vs.}\ production), etc. While this is a genuine problem, it is a problem for engineering education in general ---what should we teach?--- rather than just for the proposed criterion. In any case, we can at least account for those skills all engineers need (see for instance \citet{Markes-EJEE-06}).}
Regarding criteria~1 and~2 of \citet{Felder-tests}, one should acknowledge that dealing with tricky problems one has never seen before is something engineers have to do. Whether assigning such problems in exams is a suitable way of teaching or testing such skills is another issue. 
(Note that relying on what would make students better engineers will naturally entail a great deal of the simpler interpretations of fairness.)

\section{Fairness and beyond}
Fairness of treatment is powerless and fairness of opportunity assumes that all talents are intrinsically of equal worth, which necessarily leads to the same desert and the same grade for all students. To escape this deadlock, `what would make students better engineers' was used as a basis for decisions. This provides one with a general framework to distinguish between competing claims: the exam should favor whatever skill would be good in an engineer. Whether one uses this criterion or another, one must recognize that some such criterion is \emph{necessary} to allow the instructor to make rational decisions regarding competing claims: without such a criterion, examinations are arbitrary. The heart of the problem is that fairness is construed as the comparison of students with one another, but it does not say anything about the criteria to be used for this comparison. 

Fairness is seen as the most important criterion for exams~--- for example, students may complain that exams are unfair but are unlikely to complain that an exam does not efficiently test the knowledge and skills it is supposed to assess. One reason for this is that fairness is seen as a moral requirement whereas validity and reliability are seen as (merely) technical issues. Whether this can \emph{a priori} give fairness a special status is debatable~\cite{Bouville-misconduct}. Looking for fair exams, I found a criterion which seems both necessary for exams to be fair and not a matter of fairness. A purely relative concept such as fairness is unable to lead to concrete decisions: an external reference such as better engineers had to be introduced. Fairness is then the way this external criterion is applied: it must be the same for all students~--- this is not a direct comparison of students. Moreover, this means that fairness is a necessary consequence of the validity of the exams, rather than a separate criterion.
\hspace{\fill}\mbox{\copyright~Mathieu Bouville, 2008}

\end{document}